\documentclass[9pt,twocolumn,twoside]{opticajnl}
\journal{opticajournal} 

\setboolean{shortarticle}{false}

\usepackage{lineno}

\begin{document}

\title{Symmetric Second-Harmonic Generation in Sub-wavelength Periodically Poled Thin Film Lithium Niobate}

\author[1]{Fengyan Yang}
\author[1]{Juanjuan Lu}
\author[1]{Mohan Shen}
\author[1]{Guangcanlan Yang}
\author[1,*]{Hong X. Tang}

\affil[1]{Department of Electrical Engineering, Yale University, New Haven, CT 06511, USA}

\affil[*]{hong.tang@yale.edu}

\begin{abstract}
Second harmonic generation (SHG) extensively employs periodically poled nonlinear crystals through forward quasi-phase-matching to achieve efficient frequency conversion. As poling periods approach sub-micrometers, backward quasi-phase-matching has also been demonstrated, albeit by utilizing pulsed laser drives. The realization of symmetric second harmonic generation, characterized by counterpropagating pumps, however, has remained elusive despite theoretical predictions. The main challenge lies in achieving strong nonlinear coupling with poling period below half the wavelength of the second-harmonic light. The recent emergence of high-quality ferroelectric lithium niobate thin films provides an opportunity for achieving precise domain control at submicron dimensions. In this article, we demonstrate reliable control of ferroelectric domains in thin film lithium niobate waveguide with a poling period down to 370\,nm, thereby realizing highly efficient continuous-wave pumped symmetric SHG. This demonstration not only validates the feasibility of achieving subwavelength periodic poling on waveguides but also opens new avenues for leveraging submicron ferroelectric domain structures in integrated photonics and nonlinear optics research.  

\end{abstract}

\setboolean{displaycopyright}{false} 
\maketitle

\section{Introduction}   

As a key process in nonlinear optics, second-harmonic generation (SHG) is not only central to fundamental nonlinear science\cite{pokelcomb_alex,Lutwocolor,wjq_five_wave,comb_SHG,comb_SHG2,Cui2022_cascaded,ming_enhancedSHG}, but also 
finds applications across various fields such as precision metrology, optical clocks and quantum information processing \cite{optical_synthesizer,Papp_clock, Entangle_PRL, Lusinglephoton,luxiyuan_photonpair,amir2023phase_sensor}. It has been vastly investigated recent years on different integrated photonic platforms, such as lithium niobate (LN) \cite{wangcheng_LNSHG, Lu250000,YP_LN_poledring, Ayed_blue, LNphotonics_review}, silicon nitride (SiN) \cite{Levy2011SiN_SHG_breaksymmetry,  Timurdogan2017SiN_Efield_inducedSHG, Lu2021_SIN_photoninducedSHG}, gallium nitride (GaN) \cite{GaN_roland,XiongGaN}  and aluminum nitride (AlN) \cite{Liuxianwen23,GX2500,Alex17000}. Among these lithium niobate (LN) is particularly  noteworthy for SHG due to its large $\chi^{(2)}$ coefficients and domain engineering flexibility, which enables quasi-phase-matching (QPM) for efficient frequency conversion \cite{QPM1962PERsHAN,fejer1992QPM,boes2023LNreview,zhureviewlN}. Notably, forward second-harmonic generation (FSHG), where pump and second-harmonic waves propagate in the same direction, has been achieved in both LN straight waveguides and resonators with poling periods of several micrometers \cite{LNphotonics_review,wangcheng_LNSHG, luo2018highly,linran_adaptive_poing, Lu250000}. 

As poling periods decrease to sub-micrometers, two types of backward second-harmonic generation (BSHG) with first order QPM can be enabled 
\cite{gu1998_BSHG2, ding1996_BSHG_theory}. The first-type BSHG, primarily achieved with bulk KTP and LN crystals, produces second-harmonic (SH) light in the opposite direction as the pump input \cite{BSHG_KTP,gu1999BSHG_KTP, kang1997_BSHG,canalias2023, BSHG_LN,CW_BSHG_LN_HO}. The second-type BSHG, which involves counter-propagating pump inputs and SH outputs, however, had never been experimentally demonstrated prior to our work. Here we term this second type BSHG as symmetric SHG (SSHG) to distinguish it from the first type and FSHG. It should be noted that due to the use of bulk crystals, all the BSHG demonstrations so far utilize intense pulse lasers as pump sources. In order to enable either type of BSHGs with CW light input, significantly higher optical nonlinear coupling is required among the propagating beams and can be achieved by utilizing integrated waveguide platforms which offer tight optical confinement.
 
Despite recent advancements in achieving sub-wavelength poling in LN thin films, incorporating these techniques into the fabrication of photonic integrated circuits remains a challenge. Traditional electric field assisted poling utilizes bipolar preconditioning pulses to lower the coercive field, achieving a poling period of 737\,nm on x-cut thin film LN \cite{LNsubpole_bipolar}. Alternatively, a biased conducting tip of a scanning probe microscope allows for discretely switched domains, achieving poling periods as small as 200\,nm \cite{LN_AFMpoling, LN_AFMpoling2}. Similarly, focused ion beams have been employed for this purpose \cite{FIB,arxiv_FIB}. While these latter two approaches yield smaller periods and improved duty cycles, they are less efficient and face challenges in creating scalable poling patterns that align with the requirements of photonic circuits.

In this letter, we report a significant advancement in poling techniques, resulting in the creation of periodic poled z-cut lithium niobate nanowaveguides with a period as low as 370\,nm, with volumetric duty cycle of 37$\pm 3\%$. This breakthrough enables the first demonstration of SSHG, where first-order QPM is attained for efficient frequency doubling of a CW pump. Our 3D electrostatic simulations delineate the specific challenges arising from the E-field distribution when poling at such a small period, providing guidance for the fabrication process. The efficiency for SSHG is measured to be 1470$\%/\text{W}/\text{cm}^2$ on a waveguide with a 6\,mm-long poling section, which is comparable to that of FSHG. Notably, its phase-matching bandwidth is 250\,pm, significantly narrower than FSHG due to its special phase-matching condition, making it suitable for specific applications such as nonlinear frequency filter. This successful demonstration not only confirms the feasibility of achieving sub-wavelength poling of low-loss LN waveguide but also leads a pathway for realizing efficient backward parametric down-conversion photon sources and CW pumped mirrorless optical parametric oscillators \cite{MOPO_nspulse, mirrorlessOPO_NP, backOPO_1966,backOPO_2022, backOPO_APL,BSPDC_canalias} which are of fundamental importance for nonlinear optics and quantum information processing.

\begin{figure*}[h]
\centering
\includegraphics[width=0.9\linewidth]{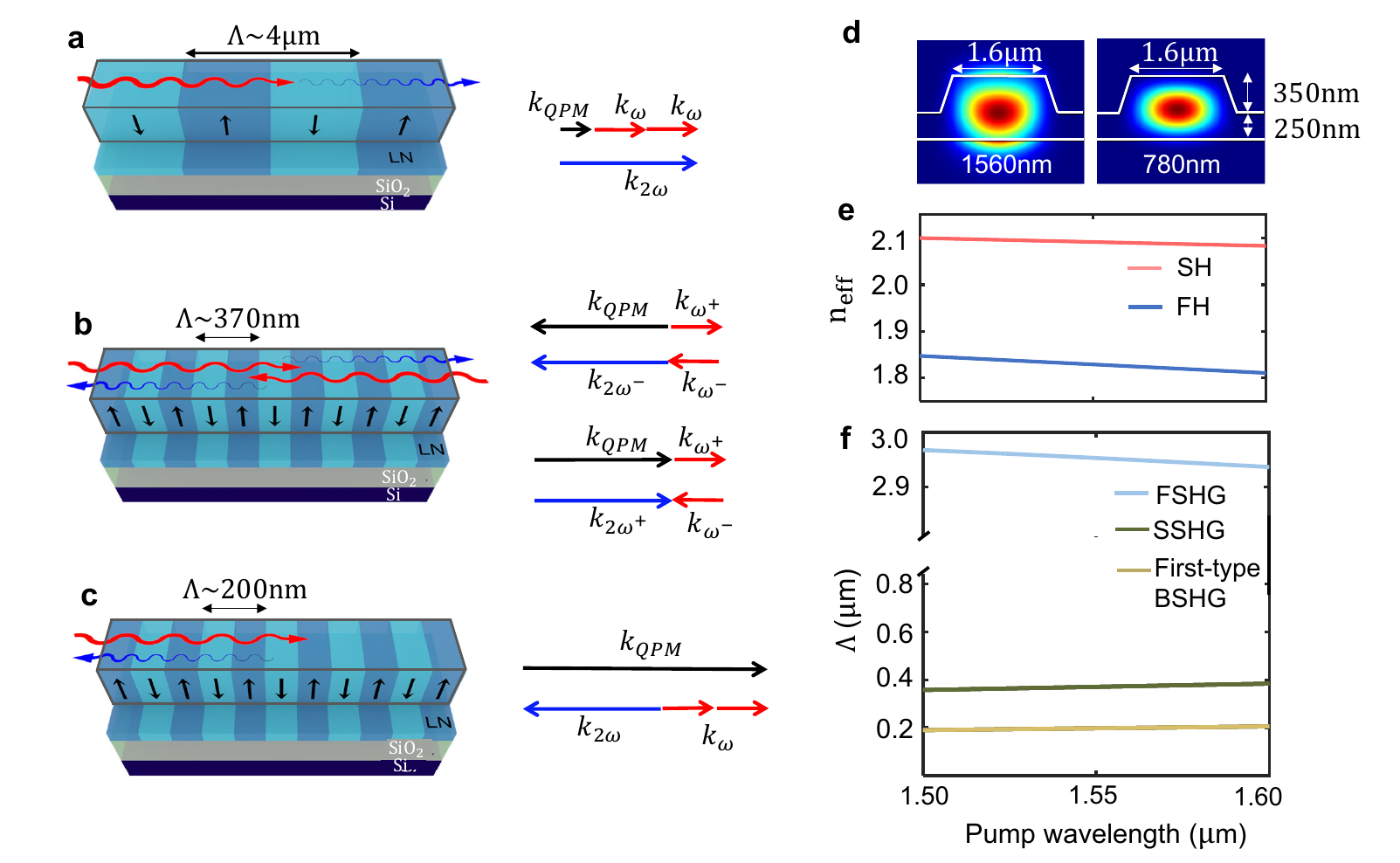}
\caption{Phase matching configuration for (a) FSHG, (b) SSHG and (c) the first-type BSHG. (d) TM$_{00}$ mode profiles at the targeted first-harmonic (FH) and second-harmonic (SH) wavelengths. (e) effective index for the modes in (d). (f) poling period for FSHG, SSHG and the first-type BSHG.}
\label{fig1}
\end{figure*}

\section{Theoretical Study of Symmetric Second Harmonic Generation}

Figure\,\ref{fig1}\,(a) shows the conventional FSHG configuration, where pump is injected from the left side of a periodically poled lithium niobate (PPLN) waveguide and the second-harmonic signal outputs from the right. The quasi-phase-matching condition for FSHG is $k_{\mathrm{QPM}}=k_{2\omega}-2k_\omega$, resulting in the poling period $\Lambda$ for first-order QPM to be several micrometers [Fig.\,\ref{fig1}\,(f)]. When poling period is further reduced to submicron level, a novel type of QPM can be attained, as shown in Fig.\,\ref{fig1}\,(b).  Here the wave vectors of the two counter-propagating pumps mutually cancel with each other, necessitating a relatively large grating vector to achieve QPM according to $k_{\mathrm{QPM}}=k_{2\omega}+k_{\omega}-k_{\omega}=k_{2\omega}$. To fulfill this condition, the typical poling period is around 370nm for conversion between pump quasi-transverse magnetic (TM) mode at 1560\,nm and second-harmonic quasi-transverse magnetic (TM) mode at 780\,nm, as indicated in Fig.\,\ref{fig1}\,(d). The first-type BSHG requires even more stringent QPM which results in poling period to be approximately 200\,nm , as shown in Fig.\,\ref{fig1}\,(c) and Fig.\,\ref{fig1}\,(f). (See supplement 1, section 2 for detailed comparison between SSHG and the first-type BSHG).

Starting from the coupled mode equations describing nonlinear interaction between the pump and SH waves, we can derive the conversion efficiency of SSHG, which under lossless condition and undepleted pump approximation is given by
\begin{equation} \label{eq:eta_lossless}
    \eta_{\mathrm{lossless}}	=	\frac{8d_{33}^2}{n_{\omega}^2 n_{2\omega} \epsilon_0 c \lambda_{2\omega}^2} \cdot \frac{S_{2\omega}}{S_\omega ^2}\cdot \mathrm{sinc}^2(\frac{\Delta k L}{2}),
\end{equation}
where $d_{33}$ is the second-order nonlinear coefficient, $n_{\omega,2\omega}$\,,  $S_{\omega,2\omega}$ are the effective refractive index and mode area for pump and SH modes, $L$ is the length of waveguide poling section and $\Delta k=k_{2\omega}-k_{\mathrm{QPM}}$ denotes the phase mismatching vector (see supplement 1, section 1). We note that this formula is consistent with that of the typical FSHG case \cite{wangcheng_LNSHG}, but the crucial difference arises from the phase mismatch vector $\Delta k$. This element plays a critical role in converting the phase matching bandwidth, defined by $\Delta k L/2=\pi$, into a scale based on wavelength ($\Delta \lambda_\omega$), thus leading to a notable distinction between the two. Specifically, for SSHG, $\Delta \lambda_\omega$ is given by	 $\lambda_\omega^2/2  L n_{\mathrm{g},2\omega}$, whereas for the ordinary SHG, it is $\Delta \lambda_\omega	= \lambda_\omega^2/2 L (n_{\mathrm{g},\omega}-n_{\mathrm{g},2\omega})$. For the structure illustrated in Fig.\,\ref{fig1}\,(d), the latter is approximately forty times larger. It's possible to further reduce the phase matching bandwidth of SSHG by engineering $n_{\mathrm{g},2\omega}$ or $L$ to be larger, thereby contributing to a narrower bandwidth in spontaneous down conversion for certain quantum photonics applications \cite{luo2020counterSPDC}.

\section{Design and Fabrication of Lithium Niobate Waveguide with Sub-wavelength Poling Period}

Domain reversal of lithium niobate is driven by the applied external electric field E, which, in conjunction with the internal depolarizing field E$_{\text{dep}}$ and screening field E$_\text{sr}$, determine the nucleation probability of reversed domains \cite{shur1999domain}. The latter two factor are material and geometry-specific, and under certain circumstances, they can even cancel each other out. Therefore, we perform 3D electrostatics simulation of the external electric field distribution using COMSOL to assess the probability of domain reversal as the poling period changes. 

\begin{figure*}[htbp]
  \centering
  \includegraphics[trim={0cm 0cm 0cm 0cm},clip,width=\linewidth]{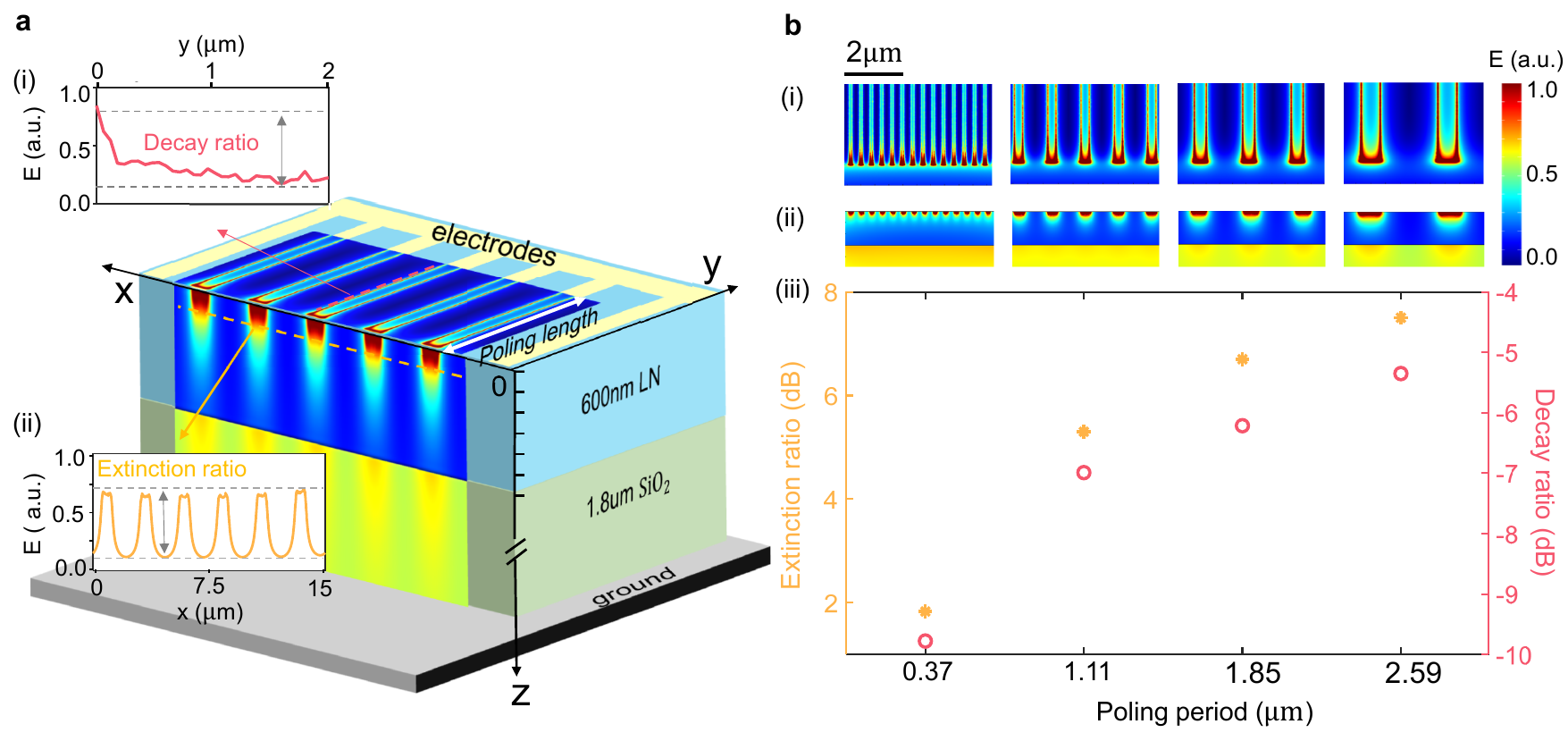}
\caption{ (a) Schematic of the wafer structure with simulated electric field distribution overlaid on the LN top surface and in a cross-section cut at the end of the electrodes. Panels (i) and (ii) depict the E-field along the red and orange dashed lines, defining the decay ratio and extinction ratio of the E-field. Data presented are for a 2.59\,{\textmu}m electrode spacing. (b) Extinction ratio (orange stars) and decay ratio (red circles) of the E-field (iii) with different poling periods for the 1st, 3rd, 5th, 7th-order QPM conditions. Corresponding top and sided E-field profiles are displayed in (i) and (ii). }
\label{fig2}
\end{figure*}

A voltage of 650V, similar to the previously reported experimental value \cite{Lu250000}, is applied to the electrode array, with the bottom Si acting as the ground. The electrode layout, shown by yellow in Fig.\,2(a), is designed to create domain patterns that harvest the largest $\chi^{(2)}$ nonlinear coefficient $d_{33}$ ($\sim $19.5\,pm/V\,\cite{LNd33}) for SHG conversion from the TM$_{00}$ pump mode to the TM$_{00}$ SH mode. Figure \ref{fig2} (a) displays the top-view and cross-sectional E-field distribution at the end of the poling electrodes with a period of 2.59\,{\textmu}m and duty cycle of 25\% for illustration purposes. To evaluate how the electrode spacing affects poling quality, we analyze the corresponding E-field features at the top surface and the cross-section with fixed electrode duty cycle, respectively.

First, nucleation tends to initiate at the electrode edges due to the large fringing field present there \cite{lang2007kinetics,QPM_domain_broaden}. Therefore, we extract the E-field along the edge of an electrode and characterize its longitudinal decay rate as $\gamma$\,=\,10log$_\mathrm{10}(E_{y=2{\mu}\text{m}} /E_{y=0})$, as shown in Fig.\,\ref{fig2}\,(a-i). 2\,{\textmu}m is the distance from the electrode end located at $y=0$, and it's chosen to be comparable to typical waveguide width. $\gamma$ decreases rapidly as the poling period becomes smaller and reaches -10\,dB at our targeted 370\,nm period, as illustrated by red circles in Fig.\,\ref{fig2}\,(b-iii). This indicates that the achievable poling length along the electrode, shown by white arrow on top surface in Fig.\,2(a), is significantly reduced and requires more precise alignment between the waveguide and poling electrodes in device fabrication. Second, a high extinction fringing field between neighboring electrodes is desirable to preserve the original domain between them. The E-field along the orange line, situated 100\,nm below the LN surface, is plotted in Fig.\,\ref{fig2} (a-ii). 
Here, the extinction ratio (ER) of the E-field is defined as $10\,\log_{10}(E_{\mathrm{max}}/E_{\mathrm{min}})$. The ER decreases significantly from 7.5\,dB to 1.8\,dB when the poling period is reduced from 2.59\,{\textmu}m to 370\,nm, as indicated by oranges stars in Fig.\,\ref{fig2} (b-iii). A reduced ER leads to excessive domain reversal beyond the designated electrode coverage area and increases the risk of overpoling. Consequently, a considerable reduction in electrode width is required to alleviate this effect for small poling periods.

\begin{figure*}[htbp]
\centering
\includegraphics[trim={0cm 0cm 0cm 0cm},clip,width=0.85\linewidth]{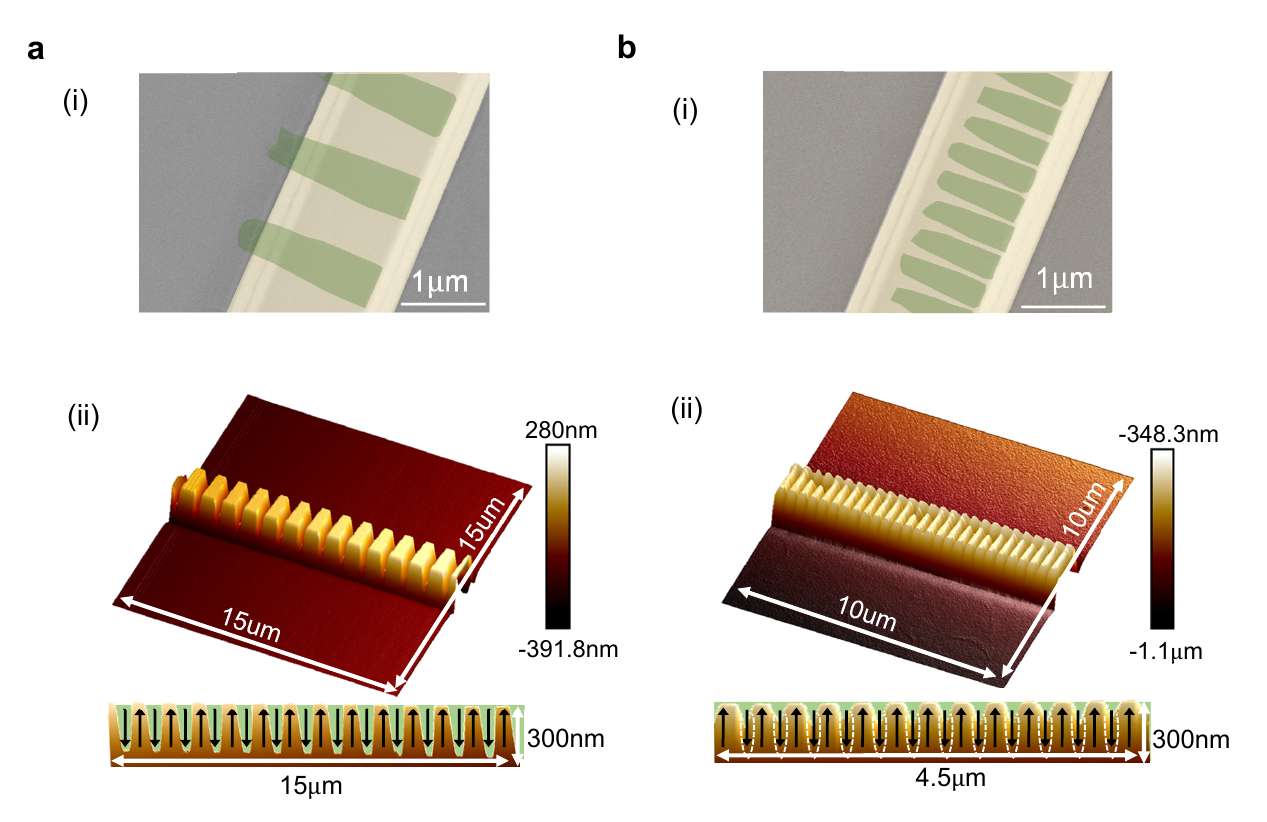}
\caption{ (a) (i) False color SEM image of a poled waveguide with $\Lambda$\,=1.11\,{\textmu}m, green regimes mark the reversed domains.  (ii) AFM mapping of waveguide surface following HF etching of the reversed domains. The corresponding side view is shown at the bottom of the panel, with domain polarization indicated by arrows.  (b) False color SEM image of a poled waveguide with $\Lambda$\,=370nm (i) and corresponding AFM mapping (ii). }
\label{fig3}
\end{figure*}

Our fabrication starts with a 600\,nm z-cut LNOI wafer. We first deposit 70\,nm nickel electrodes through a liftoff process, with the electrode width being 15$\%$ of the poling period. Such a dense electrode array requires careful selection of ebeam resist and precise liftoff control. We use MMA/PMMA bilayer ebeam resist, followed by the 3:1 iced IPA:water developer for better resolution and gentle oxygen plasma descum to remove residual resist in the patterned area. Subsequently, the nickel electrode is deposited via the ebeam evaporator and liftoff in a 80\,$^\circ$C acetone bath aided by sonication. Upon completion of the electrode fabrication, the LN film undergoes poling through a 650\,V voltage pulse lasting 10\,ms at 250\,$^\circ$C. Finally, we pattern nanowaveguides in the periodically poled LN thin film using the same process described in our previous work \cite{Lu250000}. Figure.\,\ref{fig3}\,(a-i) and (b-i) present the false color scanning electron microscopy (SEM) images of waveguides with poling period being 1.11\,{\textmu}m and 370\,nm, which are designed for third-order and first-order QPM, respectively.

To examine the switched domains, we immerse the chip in Hydrofluoric (HF) acid for 3\,mins. This duration is carefully chosen to selectively etch a portion of the reversed domains while preserving the waveguide's adherence to the wafer for subsequent domain characterization. Then we use atomic force microscopy (AFM) to map the domain distribution for two poling periods, 1.11\,{\textmu}m and 370\,nm. Fig.\,\ref{fig3}\,(a-ii) and (b-ii) present 3D AFM images for each case, along with cross-sectional views. The duty cycle for the 1.11\,{\textmu}m period is approximately 50$\%$ as observed from the side view. However, limited HF etching time on the waveguide does not adequately expose the domain profile for the 370\,nm period as excessive HF etching of the waveguides leads to their delamination from the substrate. To address this issue, we pole a separate chip consisting of only bare lithium niobate thin film under the same poling process and then apply HF etching for 20 minutes to completely removes the reversed domains without leading to the film detachment. Through this approach, the complete domain profile is exposed and mapped out by SEM and AFM and the volume duty cyle is calculated to be 37$\pm3\%$ (see supplement 1, section 3). The poling length along electrodes is measured to be 1.4\,um , which is attributed to the large E-field decay ratio as illustrated by the simulation shown in Fig.\,\ref{fig2}(b).

\section{Measurement of Symmetric  Second Harmonic Generation}
A tunable CW telecom laser (Santec 710) is split equally by a 50$\%$-$50\%$ coupler and used as pump laser. Each branch of the pump light enters the 1550\,nm port of a wavelength-division multiplexer (WDM) and is subsequently coupled into the chip via lensed fibers. SSHG signals propagate in both left (LSHG) and right (RSHG) directions, exiting the WDM through its 780\,nm port and being collected by two identical visible photon detectors. The fiber-to-chip insertion loss is calibrated to be 7.0$\pm$0.2 dB for infrared and 6.7$\pm$0.2dB for near visible. 

\begin{figure*}[h]
\centering
\includegraphics[width=\linewidth]{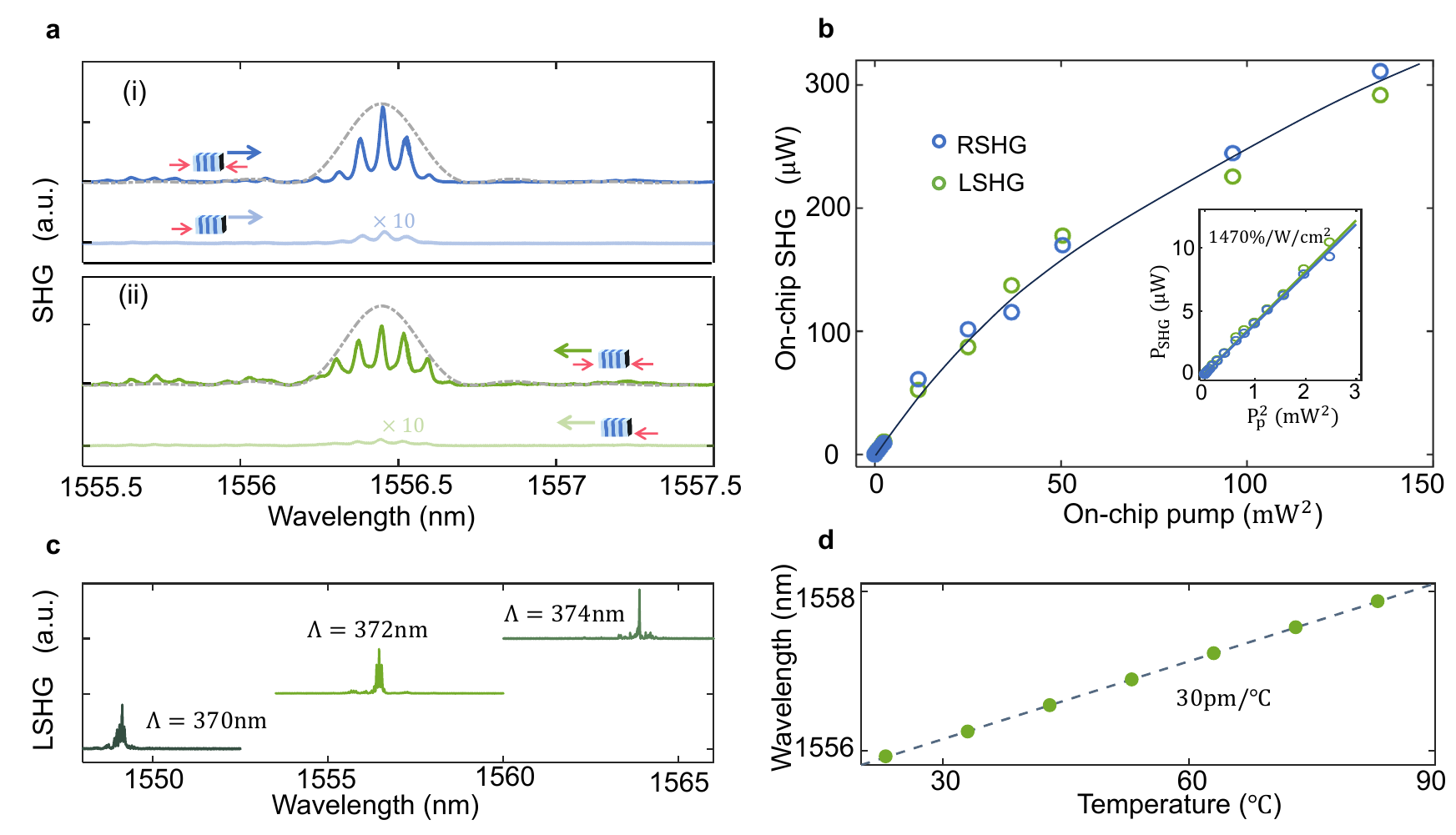}
\caption { (a) The rightward- (i) and leftward-propagating (ii) SHGs plotted in linear scale with one-side (lighter) and two-side (darker) pump inputs. The lighter curve is shifted downward vertically for clarity. The theoretical spectra are overlaid by dashed gray lines. (b) On-chip SSHG vs on-chip pump power. The inset magnifies the data in the undepleted region and presents the quadratic fitting results. (c) The shift of LSHG phase matching wavelength at varying poling periods. (d) The temperature dependence of SSHG phase matching wavelength. }
\label{fig4}
\end{figure*}

We first inject the pump laser only from the left or only from the right side of an LN waveguide with 2\,mm poling section. 
In this configuration, very weak SHG signals with similar magnitude are detected in both directions. For clarity, we only plot one of the SHG outputs, as shown by the lighter blue trace in Fig.\,\ref{fig4}\,(a-i) and the lighter green curve in Fig.\,\ref{fig4}\,(a-ii). This tiny amount of SH signal can be attributed to the pump back reflection at the waveguide facet. Subsequently, we activate the pump from both sides, resulting in two orders of magnitude higher SHG powers compared to the single-sided pump scenario. This observation confirms that the phase matching condition is satisfied only when two counter-propagating pump modes are present, i.e. $k_{2\omega}=k_{\omega}-k_{\omega}+k_{\mathrm{QPM}}$. Under the assumption of uniform phase-matching throughout the waveguide and a symmetrical measurement setup, the spectra in both directions are supposed to be identical. Indeed, as depicted in Fig.\,\ref{fig4}\,(a), the LSHG and RSHG are detected at the same pump wavelength with comparable power. The phase matching bandwidth is measured to be around 250\,pm for both LSHG and RSHG, which matches well with the theoretical spectrum (gray curve) under the undepleted pump approximation. The oscillations within the main phase matching peak are attributed to the interference pattern of a Fabry-Perot cavity formed by the waveguide facets, whose cavity length of 14\,mm gives rise to an pump FSR of $\sim$\,0.07\,nm. 


We delve further into the efficiency of SSHG in a waveguide with a 6\,mm poling section. As depicted in Fig.\,\ref{fig4}\,(b), the output powers for both left and right SHG increase quadratically with increasing pump power in the undepleted regime. The experimental normalized peak SHG efficiency $P_\text{SHG}/(P_\text{p}^2\cdot L^2)$ is found to be around 1470$\,\pm\,210
\%$\,/W/$\text{cm}^2$ for both LSHG and RSHG, where $P_\text{p}$ , $P_\text{SHG}$ are one-sided pump and SHG power. Even though the achieved effciency is lower than the theoretical prediction 2050$\,\%$\,/W/$\text{cm}^2$ due to waveguide-loss and deviation from the 50$\%$ duty cycle, the result still falls within the same order of magnitude with state-of-art efficiency in ordinary forward SHG in LN waveguide \cite{wangcheng_LNSHG,Ayed_blue} 

It is notable that the phase matching wavelength of SSHG strongly depends on the poling period according to $\lambda_\omega=2\Lambda n_{2\omega}$. In contrast, the phase matching wavelength for FSHG, given by $\lambda_\omega=2\Lambda (n_{2\omega}-n_{\omega})$, which is far less sensitive to the poling period period changes (by a factor of $\frac{n_{2\omega}}{n_{2\omega}-n_{\omega}}$). As shown in Fig.~\ref{fig4}(c), a 2\,nm variation in the poling period results in a shift of 7.32\,nm in the phase matching wavelength. This observation demonstrates that our technique can achieve nanometer-distinguishable poling periods and maintain excellent periodicity across several millimeters.
Another noteworthy result is the suppressed temperature dependence of phase-matching wavelength of SSHG, as displayed in Fig.~\ref{fig4}(d). In accordance with $\frac{d\lambda_\omega}{dT}=\frac{2\Lambda dn_{2\omega}/dT+2n_{2\omega}d\Lambda/dT}{1-2\Lambda dn_{2\omega}/d\lambda}$, the thermal shifting rate of $\lambda_\omega$ is estimated to be 40\,pm$/^\circ$C \cite{thermal_coef} and experimentally measured to be 30\,pm$/^\circ$C, whereas for FSHG, $\frac{d\lambda_\omega}{dT}=\frac{2\Lambda d(n_{2\omega}-n_\omega)/dT+2(n_{2\omega}-n_\omega)d\Lambda/dT}{1-2\Lambda d(n_{2\omega}-n_\omega)/d\lambda}$  and it's estimated to be 107\,pm$/^\circ$C mainly due to the larger poling period.


\section{Conclusion}
In conclusion, we experimentally demonstrate  symmetric  second-harmonic generation in a sub-wavelength periodically-poled lithium niobate waveguide. Our simulations and experiments reveal that the poling period is largely limited by the 600\,nm-thick LN film. Within the thin film, the poling field distribution dictates the achievable poling length along electrodes and the duty cycle. Further reduction of the poling period is possible with thinner LN films. Notably, the conversion efficiency under CW pump, measured at 1470$\%/\text{W}/\text{cm}^2$, is on par with the state-of-the-art value for the forward SHG scheme in PPLN waveguide. The unique counter-propagating pump wave configuration of SSHG leads to a narrower phase-matching bandwidth and reduced thermal shift. 
This achievement in sub-wavelength poling of LN waveguides and the successful implementation of SSHG open new avenues for complex nonlinear optical processes and quantum networks, such as CW pumped mirrorless optical parametric oscillation and spatially separable twin-photon states. These advancements will further enrich the realm of nonlinear optics and quantum information processing.

\smallskip{}
\begin{backmatter}
\bmsection{Disclosures}
The authors declare no conflicts of interest.

\bmsection{Acknowledgments}
This project is supported in part by the National Science Foundation (NSF) through ERC Center for Quantum Networks (CQN) grant (grant no EEC-1941583) and an NSF FuSe grant (grant no 2235377). The part of the research that involves lithium niobate thin film preparation is supported by the US Department of Energy Co-design Center for Quantum Advantage (C2QA) under Contract No. DE-SC0012704. The authors would like to thank Yong Sun, Lauren Mccabe, Kelly Woods, and Michael Rooks for their assistance provided in the device fabrication. The fabrication of the devices was done at the Yale School of Engineering \& Applied Science (SEAS) Cleanroom and the Yale Institute for Nanoscience and Quantum Engineering (YINQE).

\bmsection{Supplemental document}
See Supplement 1 for supporting content. 

 \end{backmatter}

\bibliography{main}

\end{document}


\maketitle

\section{Conversion efficiency of SSHG in undepleted-pump region}

The coupling mode equations describing nonlinear interaction between pump and second-harmonic (SH) waves are \cite{ding1996_BSHG_theory}

\begin{eqnarray}
\frac{dA^{\pm}_\omega}{dz} & = & \pm i\, \Gamma_\omega(A^\mp_\omega)^*\cdot(A^+_{2\omega} +A_{2\omega}^-)\cdot e^{\pm i\Delta k z}\mp\alpha_\omega A_\omega^\pm,\label{eq:couplingBSHG1} \\  
\frac{dA^{\pm}_{2\omega}}{dz} & = & \pm i\, \Gamma_{2\omega}A^+_\omega A^-_{\omega}\cdot e^{\mp i \Delta k z} \mp \alpha_{2\omega}A_{2\omega}^\pm.
\label{eq:couplingBSHG2} 
\end{eqnarray}

\noindent where $\Gamma_\omega=\frac{\omega d_{eff}}{n_\omega c}$, $\Gamma_{2\omega}=\frac{2\omega d_{eff}}{n_{2\omega}c}$ are coupling constants, $d_{eff}$=$\frac{2}{\pi} d_{33}$ is the effective second-order nonlinear coefficient, giving that the poling duty cycle is 50$\%$. $\alpha_{\omega,2\omega}$ are waveguide loss for pump and second-harmonic modes, $\Delta k= k_{2\omega}-k_{\mathrm{QPM}}$ is the phase mismatch factor. $A^\pm_\omega$, $A^\pm_{2\omega}$ refers to the amplitudes of electric field defined by

\begin{eqnarray}
E^\pm_\omega & = & A^\pm_\omega e^{i (\pm k_\omega z-\omega t)}+ c.c ,\label{eq:Efield}    \\
E^\pm_{2\omega} & = & A^\pm_{2\omega} e^{i (\pm k_{2\omega} z-\omega t)}+c.c.\label{eq:Efield2w}
\end{eqnarray}

Assuming undepleted pump and only with linear loss: 
\begin{eqnarray}
A^+_\omega(z) = e^{-\alpha_\omega( l+z)}A_0,  \label{eq:undepleted pump1} \\
A^-_\omega(z) = e^{-\alpha_\omega( l-z)}A_0,
\label{eq:undepleted pump} 
\end{eqnarray}

\noindent where  $l$ is half of the waveguide length, $A_0$ is the pump amplitude which is assumed to be equal for both propagating directions and set by pump power $P_\omega$ by$A_0=\sqrt{P_\omega/2n_\omega\varepsilon_0 c S_{\omega}}$, and $S_\omega$ is the pump mode area. Substituting Eq.\,(\ref{eq:undepleted pump1}-\ref{eq:undepleted pump}) for pump into Eq.\,(\ref{eq:couplingBSHG1}-\ref{eq:couplingBSHG2}), we obtain the equation governing the evolution of SH waves as follows,
\begin{equation}
\frac{dA_{2\omega}^\pm}{dz}=\pm i \Gamma_{2\omega}A_0^2 e^{-2\alpha_{\omega}l}\cdot e^{\mp i \Delta kz}\mp\alpha_{2\omega}A^\pm_{2\omega}.
\label{eq:SHG evolution} 
\end{equation}

This equation can be further solved by Fourier transformation with boundary conditions $A_{2\omega}^+(- l)=0$ and $A_{2\omega}^-(+l)=0$, and the result is
\begin{equation}
A^\pm_{2\omega}( z)=\frac{\beta e^{i\Delta kl-\alpha_{2\omega}l}}{i\Delta k -\alpha_{2\omega}}e^{\mp\alpha_{2\omega} z} -\frac{\beta }{i\Delta k-\alpha_{2\omega}} e^{\mp i\Delta k z},
\label{eq:A2w(z)} 
\end{equation}

where $\beta=i \Gamma_{2\omega} A_0^2 e^{-2\alpha_{2\omega}l}$. Therefore, the field amplitude square of SH waves at $z=\pm l$ can be written as

\begin{equation}
|A^+_{2\omega}( l)|^2=|A^-_{2\omega}(-l)|^2=\frac{\Gamma_{2\omega}^2|A_0|^4e^{-4\alpha_\omega l}}{\Delta k^2+\alpha_{2\omega}^2}\cdot[1+e^{-4\alpha_{2\omega}l}-2e^{-2\alpha_{2\omega}l}\cos(2\Delta k l)].
\label{eq:|A2w(l)|^2} 
\end{equation}

The conversion efficiency $\eta$ is defined by $P_{2\omega}/(P_{\omega}^2\cdot L^2)$, where $P_{\omega, 2\omega}$ are the one-sided pump and SH power, which relate to field amplitude by $P=2nc\varepsilon_0|A|^2S$, $L=2l$ is the total length of the waveguide. With the undepleted pump approximation, the conversion efficiency is given by 

\begin{equation}
\eta=\frac{P_{2\omega}}{P_\omega^2\cdot L^2}=\frac{8d_{33}^2}{n_{\omega}^2 n_{2\omega} \epsilon_0 c \lambda_{2\omega}^2 L^2} \cdot \frac{S_{2\omega}}{S_\omega ^2}\cdot \frac{e^{-2\alpha_{\omega} L}}{\Delta k^2+\alpha_{2\omega}^2}\cdot [1+e^{-2\alpha_{2\omega}L}-2e^{-\alpha_{2\omega} L}\cos(\Delta k L)].
\label{eq:conversion efficiency sup} 
\end{equation}

\noindent where $S_{\omega,{2\omega}}$ are effective mode areas for TM00 pump mode and TM00 SH mode. 
Under lossless situation, the above formula reduces to
\begin{equation} \label{eq:eta_lossless}
    \eta_{\mathrm{lossless}}	=	\frac{8d_{33}^2}{n_{\omega}^2 n_{2\omega} \epsilon_0 c \lambda_{2\omega}^2} \cdot \frac{S_{2\omega}}{S_\omega ^2}\cdot \mathrm{sinc}^2{(\frac{\Delta k L}{2})}.
\end{equation}

\section{Comparison between symmetric SSHG and the first-type backward SHG}

\begin{table}
    \centering
\caption{Comparison between symmetric SHG and the first-type backward SHG }
\label{tab1}
    \begin{tabular}{|c|c|c|} \hline 
         &  SSHG& BSHG \cite{canalias2023}\\ \hline 
         QPM&  $k_{2\omega}=k_\Lambda$& $k_{2\omega}+2k_{\omega}=k_\Lambda$\\ \hline 
          platform (poling period) 
         &  integrated LN waveguide (370\,nm) & bulk PPKTP crystal (317\,nm) \\ \hline 
         pump & CW laser & pulsed laser \\ \hline 
         efficiency&  1470$\%/$W$\cdot$ cm$^2$ & $\sim 5\times 10^{-4}\%/$W$\cdot$ cm$^2$ \tablefootnote{normalized to pulse peak power} \\ \hline 
         phase matching bandwidth&  250\,pm& 27\,pm\\ \hline 
         thermal tuning& 30\,pm/$^\circ$C & 18\,pm/$^\circ$C \\ \hline
    \end{tabular}
    
\end{table}

Table.\,\ref{tab1} presents the comparison of performance between our SSHG on integrated LNOI platform and the state-of-art fisrt-type BSHG achieved on bulk PPKTP platform \cite{canalias2023}. Both processes require stringent poling periods due to the QPM condition. Our platform demonstrates significantly higher normalized efficiency due to tight optical confinement hence allowing the use of a CW pump. The phase matching bandwidth for SSHG is usually larger, described by the expression $\frac{\lambda_\omega^2}{2  L n_{\mathrm{g},2\omega}}$, whereas for BSHG, it is given by $\frac{\lambda_\omega^2}{2 L (n_{\mathrm{g},\omega}+n_{\mathrm{g},2\omega})}$. The thermal tuning rate for SSHG is $\frac{d\lambda_\omega}{dT}=\frac{2\Lambda dn_{2\omega}/dT+2n_{2\omega}d\Lambda/dT}{1-2\Lambda dn_{2\omega}/d\lambda}$, while for BSHG, it's $\frac{d\lambda_\omega}{dT}=\frac{2\Lambda d(n_{2\omega}+n_\omega)/dT+2(n_{2\omega}+n_\omega)d\Lambda/dT}{1-2\Lambda d(n_{2\omega}+n_\omega)/d\lambda}$. This rate depends on material's thermal-optic coefficient, thermal expansion coefficient and dispersion. The experimentally reported values for both processes are on the order of $\sim$10\,pm. Achieving the first-type BSHG on our platform requires a poling period down to 200 nm according to the calculation, which is not yet feasible at this moment due to the limits imposed by the achievable contrast of the poling field.

\section{ Domain profile, volume duty cycle and efficiency estimation for 370\,nm poling period}

In the main text, we show that the reversed domain patterns can be revealed by etching the poled waveguides directly. However, for waveguides with a short poling period, such as 370\,nm, this method can not completely remove the reversed domain as extended HF etching tends to peel off the waveguides from the substrate. Here we describe our approach to accurately determine the volume duty cycle for structures poled with a 370\,nm poling period.

\begin{figure*}[h]
\centering
\includegraphics[width=0.95\linewidth]{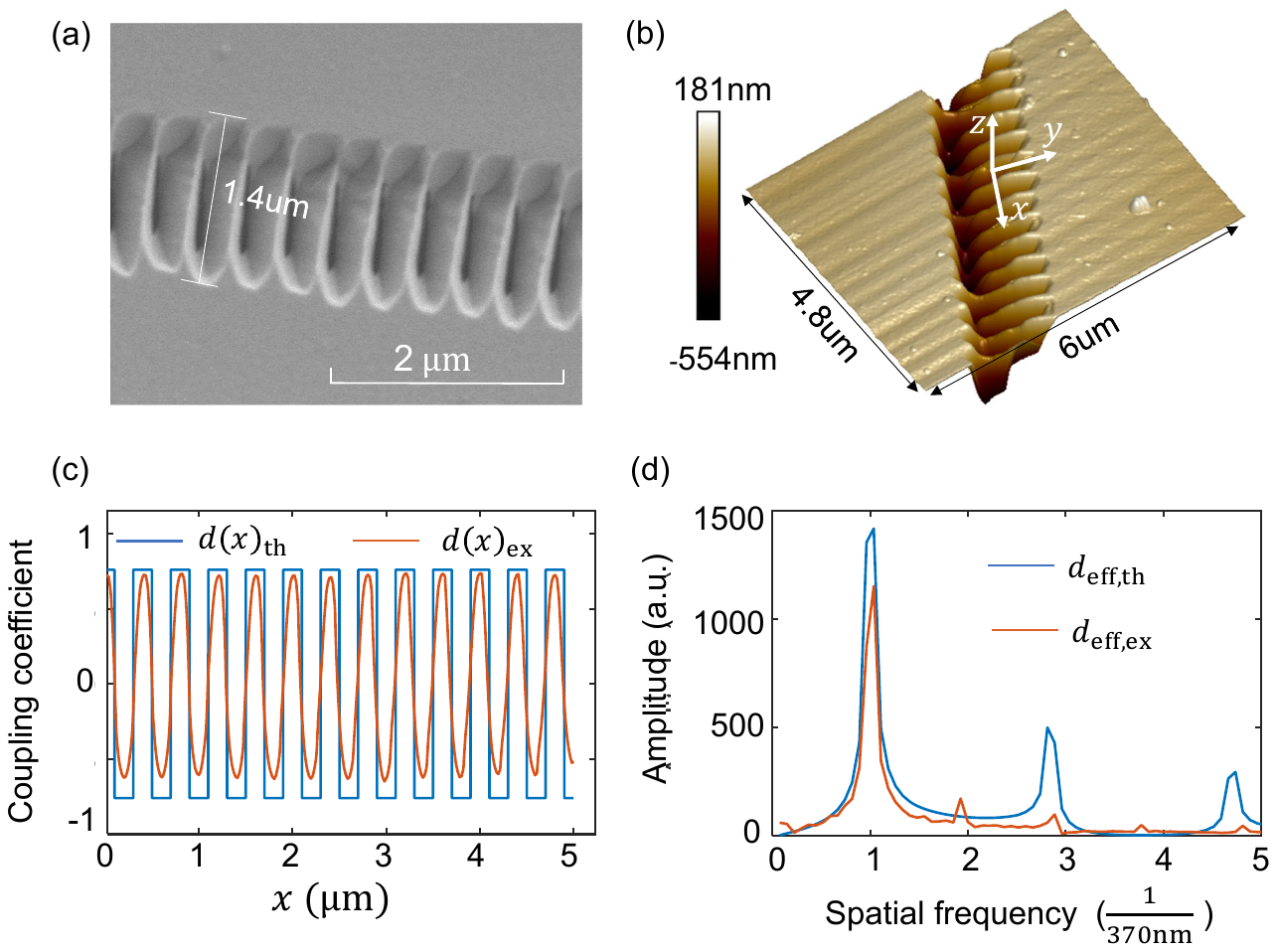}
\caption { (a) SEM image of LN thin film after HF removal of the reversed domains. (b) AFM mapping of domain profile. (c) Nonlinear coupling coefficient along the waveguide, calculated by the experimentally mapped domain profile (orange) and theoretical square-wave distribution (blue).  (d) Fourier transform of the curve in (c).}
\label{fig_sup2}
\end{figure*}

We first pole a chip consisting of thin film lithium niobate with a 370\,nm period, followed by immersion in HF solution for 20\,minutes to ensure complete etching of the reversed domains. Then scanning electron microscope (SEM) is utilized to capture the domain profiles, depicted in Fig.\,\ref{fig_sup2} (a), where the poling length is measured to be 1.4um. The 3D domain profile is further mapped out using an atomic force microscope (AFM) equipped with an ultra-sharp AFM tip (AR10-NCHR from NANOandMORE), as depicted in Fig.\,\ref{fig_sup2}(b). We export the height data of several domain profiles, integrate the surfaces to get the volume of unpoled domains, and calculate the volume duty cycle to be 37$\pm 3\%$, defined as the ratio between the poled volume and total volume.

Given the volume duty cycle $37\%$, we can estimate the upper limit of the conversion efficiency by assuming a square-wave domain distribution. This efficiency is proportional to $d_{eff}^2=(4/\pi^2)\cdot d_{33}^2 sin(D\pi)^2$ and is calculated to be 1730\,$\%$/W/$\mathrm{cm}^2$, which is higher than the experimentally measured value. The discrepancy occurs because the domain profile is not perfectly square-shaped, for example, the domain width varies along the depth direction, as shown in Fig.\,\ref{fig_sup2}(a). To obtain a more accurate estimation of $d_{eff}$ , we establish the spatial distribution of $d(x)$ by first mapping the measured height data to a $d_{33}(x,y,z)$ profile (normalized so that $d_{33}(x,y,z)$=1 or-1 for the original or reversed domain), and then integrating it with the optical mode profile according to $d(x)=\frac{\iint d_{33}(x,y,z) E_{2\omega,z}^* E_{\omega,z}^2 dydz}{\iint E_{2\omega,z}^2 dydz}$. The experimental $d(x)_{\mathrm{ex}}$  and the theoretical square-wave $d(x)_{\mathrm{th}}$, both weighted by mode distribution, are plotted in Fig.\,\ref{fig_sup2}(c). The corresponding Fourier-transformed spectrum is plotted in Fig.\,\ref{fig_sup2}(d). Since the first-order frequency component in Fig.\,\ref{fig_sup2}(d) is primarily responsible for the compensation of phase mismatch $\Delta k$, we compare $d_{\mathrm{eff, ex}}^1$ and $d_{\mathrm{eff, th}}^1$ and find their ratio is approximately 81$\%$. Through this method, we obtain a refined conversion efficiency estimation of 1350$\%$/W/$\mathrm{cm}^2$, which aligns more closely with our experimentally measured value.

\bibliography{SSHG_sup}